\documentclass[onecolumn]{aa} 
\usepackage{graphicx}
\usepackage{txfonts}
\authorrunning{Takahashi and Takahashi}
\titlerunning{Trapped photons around Kerr black hole}
\begin{document}
\title{Anisotropic radiation field and trapped photons around the Kerr black hole}

\author{
R. Takahashi\inst{1} 
\and 
M. Takahashi\inst{2}}

\offprints{R. Takahashi}

\institute{
Cosmic Radiation Laboratory, the Institute of Physical and Chemical Research, 
2-1 Hirosawa, Wako, Saitama 351-0198, Japan\\
\email{rohta@riken.jp}
\and 
Department of Physics and Astronomy, Aichi University of Education, Kariya, 
Aichi 448-8542, Japan\\
\email{takahasi@phyas.aichi-edu.ac.jp} 
}

\date{Received August XX, 20XX; accepted March XX, 20XX}
 
\abstract 
{}
{
In order to understand the anisotropic properties of local radiation field in the curved spacetime 
around a rotating black hole, we investigate the appearance of  a black hole seen by an observer 
located near the black hole. When the black hole is in front of a source of illumination the black 
hole cast shadow in the illumination. Accordingly, the appearance of the black hole is called 
the black hole shadow. 
}   
{
We first analytically describe the shape of the shadow in terms of constants of motion for a photon 
seen by the observer in the locally non-rotating reference frame (LNRF). Then, we newly derive the 
useful equation for the solid angle of the shadow. In a third step, we can easily plot the apparent image 
of the black hole shadow. Finally, we also calculate the ratio of the photon trapped by the 
hole and the escape photon to the distant region for photons emitted near the black hole. 
}
{
From the shape and the size of the black hole shadow, we can understand the signatures of the 
curved spacetime; i.e., the mass and spin of the black hole. Our equations for the solid angle of the 
shadow has technical advantages in calculating the photon trapping ratio. That is, this equation is 
computationally very easy, and gives extremely precise results. This is because this equation is 
described by the one-parameter integration with given values of the spin and location for the black 
hole considered. After this, the solid angle can be obtained without numerical calculations of the null 
geodesics for photons. 
}
{}


\maketitle
%

\section{Introduction}
To understand the general relativistic effects on the radiation field around a black hole it is
essentially important to elucidate the nature of both the physics and the astrophysics in the curved 
spacetime. This is because the radiation field plays energetically and dynamically important roles 
in the accretion flows plunging into the black hole, especially for the supercritical or  hypercritical 
accretion flows, where the photons or neutrinos interact with matter. Even for black holes
with low mass accretion rates, the general relativistic effects of the
radiation field becomes important because the observational signatures
directly reflect the nature of the radiation field. So far, the
radiative transfer in curved spacetime was investigated by many authors
(e.g. Lindquist \cite{L66}; Anderson\& Spiegel \cite{AS72}; Schmid-Burgk
\cite{S78}; Thorne \cite{T81}; Schinder \cite{S88}; Turolla \& Nobili
\cite{TN88}; Anile \& Romano \cite{AR92}; Cardall \& Mezzacappa
\cite{CM03}; Park \cite{P06}; Takahashi \cite{T07b}; \cite{T08}; De
Villiers \cite{V08}; Farris et al. \cite{F08}).

It is believed that the spacetime geometry around a black hole in the real universe is well described 
by the Kerr metric. In this case, clarifying the effects of the black hole spin on 
the radiation field is required to understand the nature of the radiation field in the black hole
spacetime. Any radiation field around the rotating black hole is 
influenced by frame-dragging effects, which make an anisotropy of the
local radiation field. That is, the anisotropic features of the local
radiation field directly reflects the effects of the black hole spin. In
the optically thick limit of the radiation field, it is expected that
the anisotropy becomes very small (but not zero), because the radiation
field and the matter intensely interact. On the other hand, in the
optically thin limit of the radiation field the anisotropic nature can
be clearly seen in the vicinity of the black hole. For example, for the
black holes in the center of the gamma-ray bursts (GRBs) the effects of
the neutrino annihilations, which are potentially the energy source of
the explosion of the GRB, are well investigated in the optically thin
limit of the neutrino radiation field (e.g. Asano \& Fukuyama
\cite{AF01}; Miller et al. \cite{M03}; Birk et al. \cite{B07}). In this
case, the black hole's rotation enhances the deposition energy due to
the neutrino annihilations. In the optically thin limit, the photons or
neutrinos emitted at some location near the black hole are transfered
along the null geodesics and some of them are swallowed by the black
hole. As is well known, because the propagation rays are influenced by the
frame-dragging effects due to the black hole's rotation, at some
location near the black hole the effects of the black hole spins are
directly seen in the shape and the size of the appearance of the black
hole. When a black hole is in front of a source of illumination, the
black hole casts a shadow on the illumination. Accordingly, the appearance of
the black hole is sometimes called a black hole shadow (e.g. Falcke,
Melia \& Agol \cite{FMA00}; Takahashi \cite{T04}, \cite{T05}; Huang et
al. \cite{H07}; Bambi \& Freese \cite{BF09}; Hioki \& Maeda
\cite{HM09}) or silhouette (e.g. Fukue \cite{F03}; 
Broderick \& Loeb \cite{BL09}; Schee \& Stuchl\'{i}k \cite{SS09a}). 
With regard to the realistic accretion flows around black holes, when
we observe the black hole with accreting gases, we only see the escape
photons emitted by surrounding gases, while the trapped photons make up the
shade region.

The observers near the black hole see the black hole shadow in their sky. In 
this paper we calculate the analytic expression of the shape of the shadow seen by the
observer located near the black hole. We also obtain the
new equation for the solid angle of the shadow in the sky of the
observer. The main purpose of the present study is to show the analytic
expressions of the shadow and the equation for the solid angle. Our
equation for the solid angle of the shadow has technical advantages. By
using this equation, the calculations of the null geodesics around the
Kerr spacetime is not required for the calculations of the solid
angle. Also, we do not need the particles to be emitted and judge which particles
will be swallowed by the black hole, as in the calculations done in
e.g. Thorne (\cite{T74}). Since the equation for the solid angle
presented in this paper is described by the one-parameter integration
for the given values of the black hole spin and the location, this
equation is computationally very easy. We assume $c=1$
and $G=1$. After giving some preliminary calculations in
Sect. \ref{sec:pre}, in Sect. \ref{sec:bhs} we give the analytic expressions for
the shape of the shadow seen by the observer near the black hole. In
Sect. \ref{sec:trap}, the equation for the solid angle of the shadow is 
presented. We give the conclusions in Sect. \ref{sec:con}.

\section{Preliminaries}
\label{sec:pre}

\subsection{Background metric}
We describe the background metric for a Kerr black hole 
by using the spherical Boyer-Lindquist coordinate.  
The 3+1 form of the metric $g_{\mu\nu}$ is described as 
\begin{eqnarray}
ds^2 =-\alpha^2dt^2 +\gamma_{ij}(dx^i+\beta^i dt)(dx^j+\beta^j dt).   
\end{eqnarray}
On the other hand,  the inverse of the metric $g^{\mu\nu}$ is given as 
\begin{eqnarray}
g^{\mu\nu}\partial_\mu \partial_\nu 
	=
	\frac{-1}{\alpha^2}(\partial_t -\beta^i\partial_i)(\partial_t -\beta^j\partial_j)+\gamma^{ij}\partial_i \partial_j. 
\end{eqnarray}
For the Kerr metric described by the spherical Boyer-Lindquist coordinate, 
the non-zero components of the lapse function ($\alpha$), the
shift vector ($\beta^i$ and $\beta_i$) and the spatial components of the
metric ($\gamma_{ij}$ and $\gamma^{ij}$) are given as  
\begin{eqnarray}
\alpha=\sqrt{\frac{\Delta\rho^2}{A}},~~~
\beta^\phi=-\omega,~~~
\beta_\phi=-\frac{2Mar\sin^2\theta}{\rho^2},~~~
\gamma_{rr}=\frac{\rho^2}{\Delta},~~~
\gamma_{\theta\theta}=\rho^2,~~~
\gamma_{\phi\phi}=\frac{A\sin^2\theta}{\rho^2},~~~
\gamma^{rr}=\frac{\Delta}{\rho^2},~~~
\gamma^{\theta\theta}=\frac{1}{\rho^2},~~~
\gamma^{\phi\phi}=\frac{\rho^2}{A\sin^2\theta},~~~
\end{eqnarray}
and $\gamma_{ij} = \gamma^{ij} = 0$ for all other $i$, $j$. 
Here, $\Delta = r^2-2Mr+a^2$, $\rho^2 = r^2 + a^2 \cos^2\theta$,
$A=(r^2+a^2)^2-a^2\Delta\sin^2\theta$ and $\omega={2Mar}/{A}$ and   
we use $\beta_i =\gamma_{ij}\beta^j$ and $\gamma^{ik}\gamma_{kj}=\delta^i_j$. 
The determinant of $\gamma_{ij}$ is given as 
${\rm det}\gamma_{ij}=\gamma_{rr}\gamma_{\theta\theta}\gamma_{\phi\phi}
	=(\gamma^{rr}\gamma^{\theta\theta}\gamma^{\phi\phi})^{-1}	
	={A\rho^2\sin^2\theta}/{\Delta}$. 

\subsection{Locally non-rotating reference frame tetrads}
The locally non-rotating reference frame (LNRF) is the frame moving with the four velocity given as $u^\mu=\alpha\delta_t^\mu$. For this frame, the tetrad vectors are given as 
\begin{eqnarray}
e^{(0)}_\alpha = \left[\alpha,~0,~0,~0\right],~~~~~
e^{(1)}_\alpha = \left[ 0,~\sqrt{\gamma_{rr}},~0,~0\right],~~~~~
e^{(2)}_\alpha = \left[ 0,~0,~\sqrt{\gamma_{\theta\theta}},~0\right],~~~~~
e^{(3)}_\alpha = \sqrt{\gamma_{\phi\phi}}\left[ \beta^\phi,~0,~0,~1 \right].~~~
\end{eqnarray}
In the same way, we have 
\begin{eqnarray}
e_{(0)}^\alpha = \frac{1}{\alpha} \left[1,~0,~0,~-\beta^\phi \right],~~~~~
e_{(1)}^\alpha = \left[0,~\sqrt{\gamma^{rr}},~0,~0\right],~~~~~
e_{(2)}^\alpha = \left[0,~0,~\sqrt{\gamma^{\theta\theta}},~0\right],~~~~~
e_{(3)}^\alpha = \left[0,~0,~0,~\sqrt{\gamma^{\phi\phi}}\right].~~~ 
\end{eqnarray}

\subsection{Photon four momentum}
The photon four momentum of the photon is calculated from the action $S$ as $p_\alpha = \partial_\alpha S$. The action $S$ is given as   
\begin{eqnarray}
S = \frac{1}{2}m^2 \lambda -E t+L_z \phi + s_r \int^r \frac{\sqrt{R}}{\Delta}dr +s_\theta \int^\theta \sqrt{\Theta} d\theta, 
\end{eqnarray}
where $\lambda$ is the affine parameter, $m$ is the mass of the particle, $E$ is the energy and $L_z$ is the angular momentum with respect to $z$-axis. The signs $s_r (=\pm 1)$ and $s_\theta (=\pm 1)$ correspond to the directions of the propagation of the null geodesics, i.e. the signs of $p_r$ and $p_\theta$ determine the signs $s_r$ and $s_\theta$. The functions $\Theta(\theta)$ and $R(r)$ are given as 
\begin{eqnarray}
\Theta(\theta) &\equiv& \mathcal{Q}-\cos^2\theta
 \left(-a^2E^2+L_z^2/\sin^2\theta\right), 
 \\ 
 R(r)&\equiv& [(r^2+a^2)E-aL_z]^2-\Delta\left[(L_z-aE)^2+\mathcal{Q}\right]. 
\end{eqnarray}
Here, $\mathcal{Q}$ is the Carter constant. Along the null geodesics, $E$, $L_z$ and $\mathcal{Q}$ are constants of motion. The photon momentum is explicitly given as 
\begin{eqnarray}
p_t =\partial_t S = -E,~~~~~
p_\phi = \partial_\phi S= L_z,~~~~~
p_r = \partial_r S 
= s_r \frac{\sqrt{R}}{\Delta},~~~~~
p_\theta = \partial_\theta S 
=s_\theta\sqrt{\Theta}. 
\end{eqnarray}
Hereafter we newly define 
\begin{eqnarray}
r_*\equiv \frac{r}{M},~~~
a_*\equiv \frac{a}{M},~~~
\Delta_*\equiv \frac{\Delta}{M^2}=r_*^2-2r_*+a_*^2,~~~
A_*\equiv \frac{A}{M^4},~~~
\omega_*\equiv \omega M,~~~
\rho_*^2\equiv \frac{\rho^2}{M^2},
\end{eqnarray}
and 
\begin{eqnarray}
\zeta \equiv \frac{L_z}{ME},~~~
\eta \equiv \frac{\mathcal{Q}}{M^2E^2},~~~
R_*(r_*) \equiv \frac{R(r)}{E^2 M^4},~~~
\Theta_*(\theta) \equiv \frac{\Theta(\theta)}{M^2E^2}, 
\end{eqnarray}
then we obtain 
\begin{eqnarray}
R_*(r_*)&=&[(r_*^2+a_*^2)-a_*\zeta]^2-\Delta_*\left[(\zeta-a_*)^2+\eta\right], 
\\ 
\Theta_*(\theta) 
	&=&\eta+(a_*-\zeta)^2-(a\sin\theta-\zeta/\sin\theta)^2. 
\end{eqnarray}
Since $\Theta$ cannot be negative, there is a constraint of 
$\eta+(a_*-\zeta)^2 \ge 0$. 
Now, for $\eta-(a_*-\zeta)^2=0$, we  have $\sin^2\theta=\zeta/a_*$; 
that is, $\theta=$constant. Hereafter, we drop the asterisk of $r_*$ for
simplicity. The photon four momentum in the LNRF is calculated as 
\begin{eqnarray}
-p^{(t)}=p_{(t)}
	=\frac{1}{\alpha} (p_t-\beta^\phi p_\phi),~~~
p^{(r)}=p_{(r)}
	= \sqrt{\gamma^{rr}} p_r,~~~
p^{(\theta)}=p_{(\theta)}
	= \sqrt{\gamma^{\theta\theta}} p_\theta,~~~
p^{(\phi)}=p_{(\phi)}
	= \sqrt{\gamma^{\phi\phi}} p_\phi.~~~
\end{eqnarray}

\subsection{Innermost unstable orbit}
In the motion of the $r$-direction, the radius of the innermost unstable orbit $r_{s}$ for 
a photon is calculated from the equations given as 
$R_*(r_{s}) = 0$ and $\partial_{r_{s}} R_*(r_{s})=0$. 
From these equations, we have 
\begin{eqnarray}
\zeta = \frac{r_s^2-a_*^2 - r_s \Delta_*(r_s)}{a_*(r_s-1)},~~~ 
\eta=\frac{r_s^3}{a_*^2(r_s-1)^2}[4\Delta_*(r_s)-r_s(r_s-1)^2]~~~{\rm for~~}a_*\neq 0,~~~~~
\eta+\zeta^2=27,~~~r=3\label{eq:zetaeta_a0}~~~{\rm for~~}a_*=0.~~~~~
\label{eq:zetaeta}
\end{eqnarray}
These results are presented in the past studies (e.g., Bardeen \cite{B73}; 
Chandrasekhar \cite{C83}). 
We note that the innermost unstable orbit was calculated in a different way  
in the recent study by Schee \& Stuchl\'{i}k (\cite{SS09a}), 
\footnote{
In Schee \& Stuchl\'{i}k (\cite{SS09a}), the parameter $\mathcal{L} (=\zeta^2+\eta)$ 
is used as one of the constants along the geodesics. By using this, the functions 
$R_*(r_*)$ and $\partial_{r_*}R_*(r_*)$ are related to $\mathcal{L}$ and 
$\partial_{r_*} \mathcal{L}$ as  
\begin{eqnarray}
\mathcal{L}
	= 2a_*\zeta-a_*^2 + \frac{(r_*^2+a_*^2-a_*\zeta)^2-R_*}{\Delta_*},~~~~~
\partial_{r_*}\mathcal{L}
	= \frac{2(r_*-1)R_*}{\Delta_*}-\partial_{r_*}R_* 
		+\frac{2(r_*^2+a_*^2-a_*\zeta)}{\Delta_*}\left[
			2r_*\Delta_*-(r_*-1)(r_*^2+a_*^2-a_*\zeta)
		\right]. ~~~~
\end{eqnarray}
By using $R_*(r_*)=0$, $\partial_{r_*}R(r_*)=0$ and $\zeta$ given in Eq. (\ref{eq:zetaeta}), 
it is shown that $\mathcal{L}_{\rm max}=0$ and  $\partial_{r_*}\mathcal{L}_{\rm max}=0$. 
Here, when $R_*(r_*)=0$, $\mathcal{L}=\mathcal{L}_{\rm max}$ 
where $\mathcal{L}_{\rm max}$ is given by Eq. (41) in Schee \& Stuchl\'{i}k (\cite{SS09a}). 
From these calculations we can see that the calculation method for the innermost unstable 
orbits used in Schee \& Stuchl\'{i}k (\cite{SS09a}) produces the same results in terms 
of $\zeta$ and $\eta$ as in the past studies of Bardeen (\cite{B73}) and 
Chandrasekhar (\cite{C83}). 
}
in which deeper explanations of the innermost unstable orbit 
are given (see Sect. 3.3 in Schee \& Stuch\'{i}k \cite{SS09a}).

\section{Shape of the black hole shadow as seen by the locally non-rotating reference frame observer}
\label{sec:bhs}
Here we calculate the shape of the black hole shadow in the sky of the LNRF observer near the black hole. In order to do this, we first  introduce the angular coordinates $(\bar{\theta},~\bar{\phi})$ in the sky of the observer. The direction of the null geodesics passing the observer is described by these coordinates. That is, the photon momentum in the LNRF can be expressed as 
\begin{eqnarray}
p^{(0)}=E,~~~~~
p^{(1)}=-E\cos\bar{\theta},~~~~~
p^{(2)}=E\sin\bar{\theta}\cos\bar{\phi},~~~~~
p^{(3)}=E\sin\bar{\theta}\sin\bar{\phi}. 
\label{eq:pLNRFthph}
\end{eqnarray}
These relations define the angular coordinates $(\bar{\theta},~\bar{\phi})$. In these definitions, 
the direction of $\bar{\theta}=0$ corresponds to the direction of the black hole. Here, we assume 
\begin{eqnarray}
[x^{(0)},~x^{(1)},~x^{(2)},~x^{(3)}]=[t,~r,~\theta,~\phi]. 
\end{eqnarray}
In this case, the direction of $p^{(r)}$ corresponds to the angle $\bar{\theta}=\pi$ and 
the direction of $p^{(\theta)}$ corresponds to $\bar{\phi}=0$. 

On the other hand, by using the LNRF tetrads given above, we have 
\begin{eqnarray}
\frac{p^{(r)}}{p^{(t)}} =\frac{s_r\sqrt{R_*/A_*}}{1-\omega_*\zeta},~~~~~
\frac{p^{(\theta)}}{p^{(t)}} 
	= \left(\frac{\Delta_*}{A_*}\right)^{1/2}\frac{s_\theta\sqrt{\Theta_*}}{1-\omega_* \zeta},~~~~~
\frac{p^{(\phi)}}{p^{(t)}} 
	=\left(\frac{\Delta_*^{1/2}\rho_*^2}{A_*\sin\theta}\right) \frac{\zeta}{1-\omega_* \zeta}.
\label{eq:pratioLNRF}
\end{eqnarray}
The parameters $(\zeta,~\eta)$ describing the shape of the black hole shadow are given by Eq. (\ref{eq:zetaeta}) (see, e.g. Bardeen \cite{B73}; Chandrasekhar \cite{C83}). From the definitions of $(\bar{\theta},~\bar{\phi})$ in Eq. (\ref{eq:pLNRFthph}), we have 
\begin{eqnarray}
\bar{\theta}=\cos^{-1}\left[-p^{(r)}/p^{(t)}\right],~~~
\bar{\phi}=\tan^{-1}\left[\frac{p^{(\phi)}/p^{(t)}}{p^{(\theta)}/p^{(t)}}\right]. 
\label{eq:thphb}
\end{eqnarray}
With these relations we can specify the position in the sky of the observer from the photon momentum and vice versa. By using Eqs. (\ref{eq:pratioLNRF}) and (\ref{eq:thphb}), we can draw the contour of the black hole shadow in the sky of the observer located at $(r, \theta)$. Along the contour of the shadow,  we change the parameter $r_{s}$ of Eq. (\ref{eq:zetaeta}) in the range of $\Theta_*\ge 0$ (see, e.g. Bardeen \cite{B73}; Chandrasekhar \cite{C83}). Hereafter, we introduce new coordinates 
\begin{eqnarray}
X=\sin\bar{\theta}\sin(-\bar{\phi}),~~~~
Y=\sin\bar{\theta}\cos(-\bar{\phi}),~~~~
Z=\cos\bar{\theta}. 
\label{eq:XYZ} 
\end{eqnarray}
With the coordinates $(X,Y,Z)$, we plot the projection of the black hole
shadow onto the sky. This contour is expressed by a closed curve in the
sphere (see Fig \ref{fig:bhsc_allsky2}; discussed below). 
Before doing this, we have to specify the signs of $s_r$ and $s_\theta$. To plot the contour of the shadow, there are two angles of $\Theta(\theta)=0$, where the sign of $s_\theta$ changes across these angles.  
When the observer is located near the black hole, there are points where the sign of $s_r$ changes. The sign of $s_\theta$ is determined by 
\begin{eqnarray}
s_\theta=+1~~~{\rm for~~~}Y>0,~~~~~s_\theta=-1~~~{\rm for ~~~}Y<0.~~~
\end{eqnarray}
On the other hand, when the observer is located far from the black hole, we have $s_r=-1$. This means that all the photons swallowed by the black hole move toward the black hole, i.e. $p^r<0$. However, when the observer is located near the black hole, even if the photons are emitted in the outgoing direction $p^r>0$, some of them are swallowed by the black hole. In this case, there are points where the sign of $s_r$ changes in the contours of the shadow. The sign of $s_r$ is determined as 
\begin{eqnarray}
s_r=+1~~~{\rm for~~~}\partial_{r_s}R(r)>0 ,~~~~~s_r=-1~~~{\rm for~~~}\partial_{r_s}R(r)<0. 
\end{eqnarray}
We note that in $R(r)$, only $\zeta$ and $\eta$ depend on the value of $r_s$. 

\begin{figure}
\begin{center}
\vspace{0mm}
\includegraphics[width=120mm, angle=0]{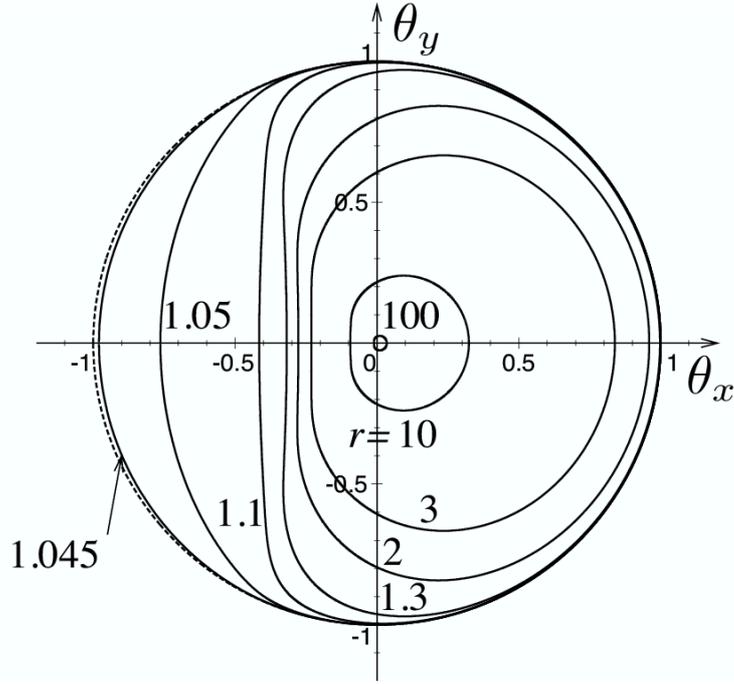}
\vspace{0mm}
\caption{\label{fig:bhsc_allskyL}
Shapes of the black hole shadow seen by the observer near the black hole. In this plot, the normalized Lambert map projection described by Eq (\ref{eq:Lambert}) is used. The abscissa is $\theta_x$ and the ordinate is $\theta_y$. In this figure, the direction of the black hole ($\bar{\theta}=\pi$) corresponds to $(\theta_x,~\theta_y)=(0,~0)$, and the opposite direction ($\bar{\theta}=0$) corresponds to the outermost circle (i.e., $\theta_x^2+\theta_y^2=1$) denoted by the dotted line. The projection of the rotation axis of a black hole corresponds to $\theta_x=0$. The locations of the observers are $\theta=85^\circ$ and $r=100,~10,~3,~2,~1.3,~1.1,~1.05,~1.045$. The black hole spin is $a_*=0.999$. In this case, the radius of the event horizon is $r_+\sim1.0447M$. 
}
\end{center}
\end{figure}

\begin{figure}
\begin{center}
\vspace{0mm}
\hspace*{0mm}
\includegraphics[width=70mm, angle=0]{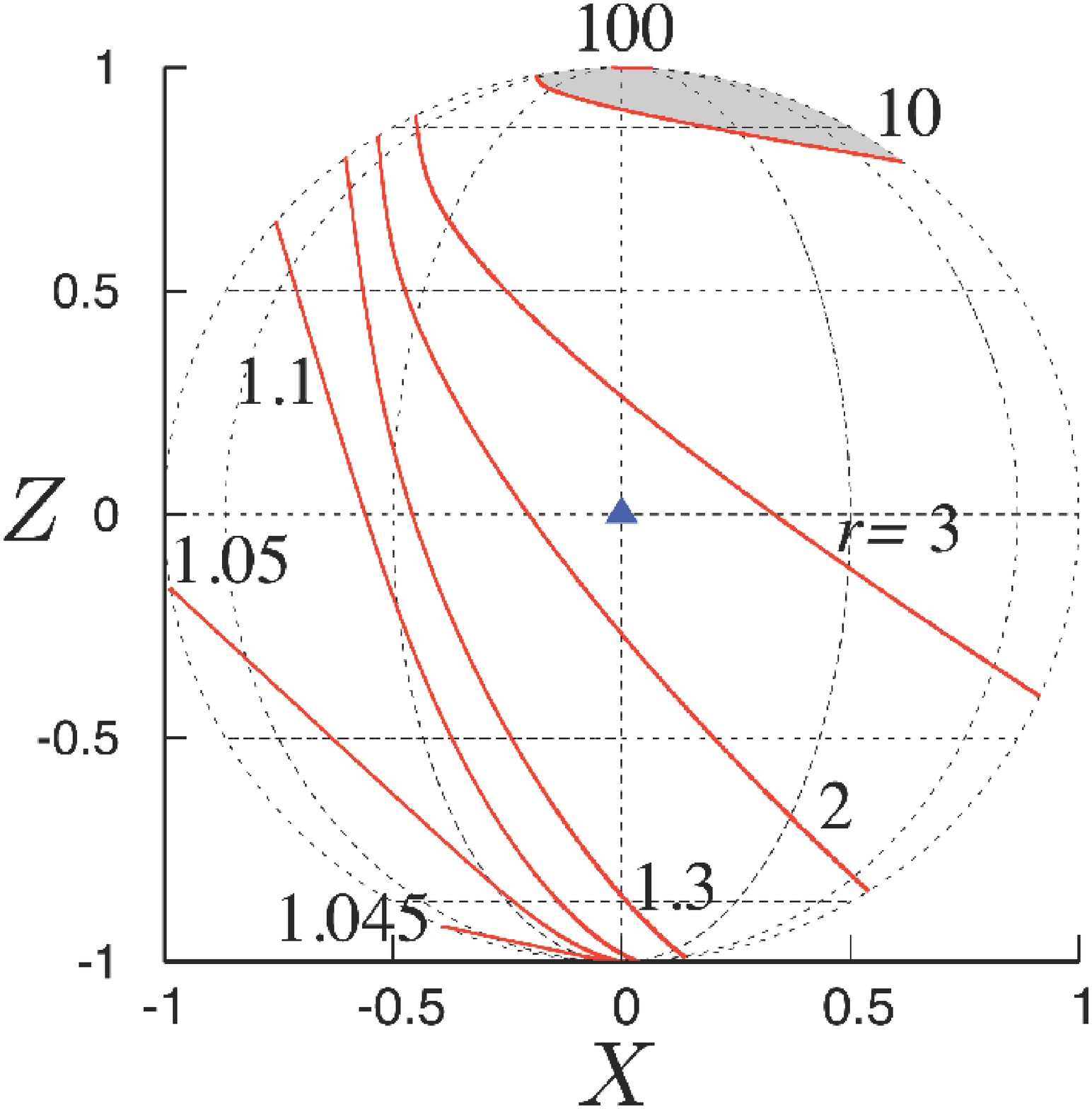}\hspace{5mm}\vspace{3mm}\\
\includegraphics[width=85mm, angle=0]{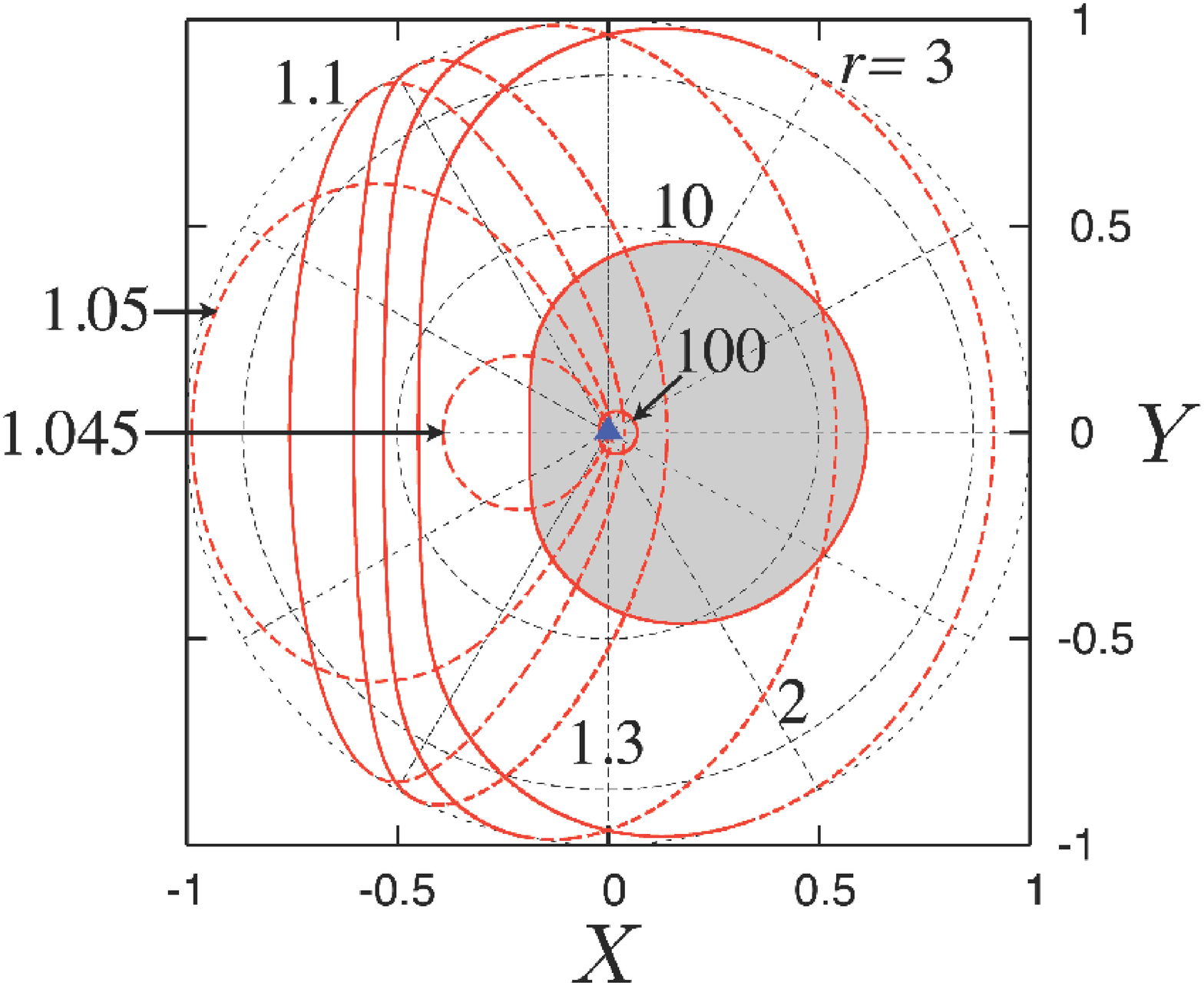}\vspace{3mm}\\
\hspace*{-10mm} \includegraphics[width=100mm, angle=0]{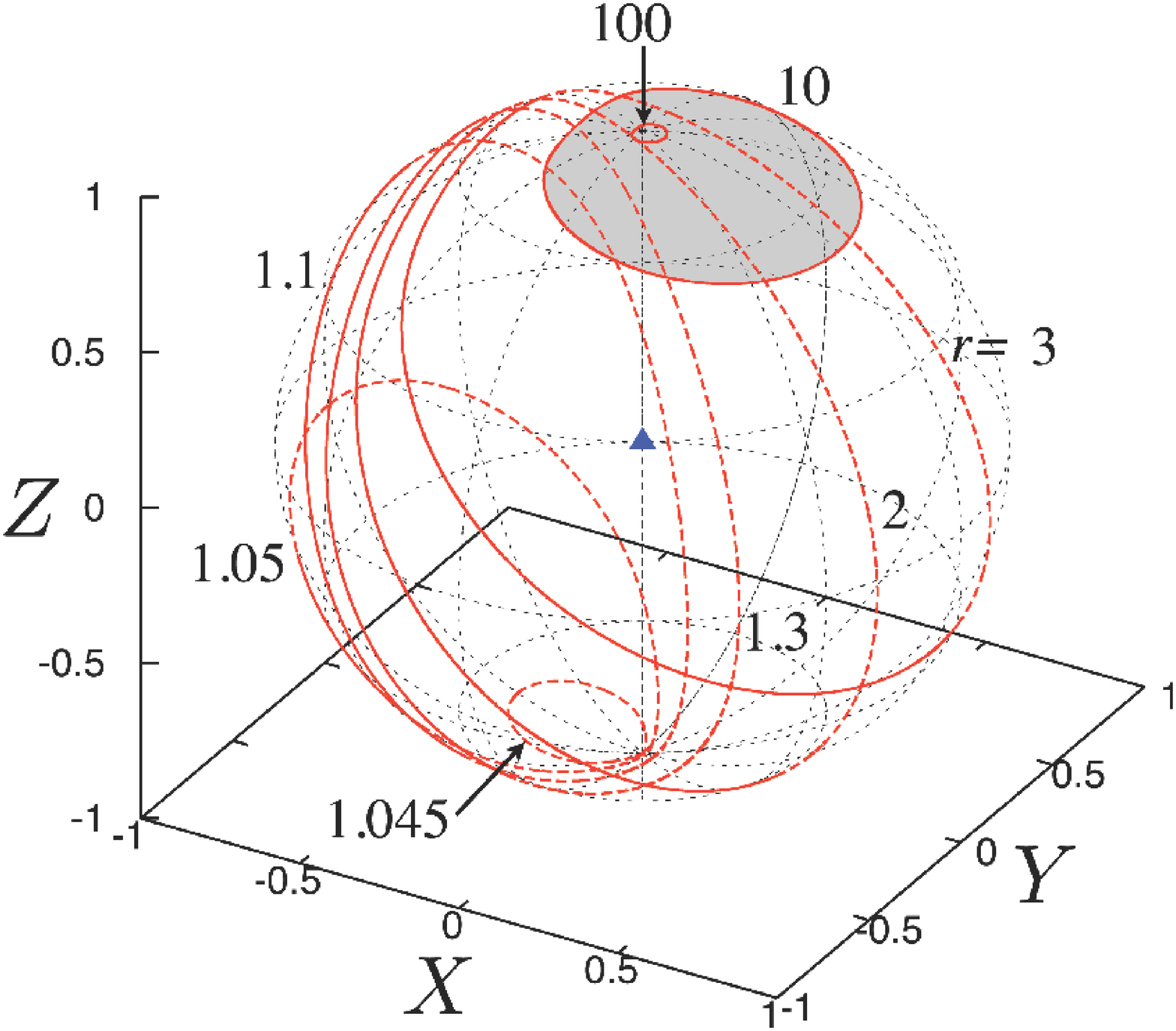}
\caption{ \label{fig:bhsc_allsky2}
Black hole shadow in the sky of the observer near the black hole in the three dimensional plot $(X,~Y,~Z)$. The parameters are the same as those in Fig. \ref{fig:bhsc_allskyL}. The position of the observer is at the origin $(X, Y, Z)=(0, 0, 0)$ ({\it filled triangles}). In the sky of the observer, the direction of the black hole is $(X, Y, Z)=(0, 0, 1)$. The shaded regions are shadows for $r=10$. The contours of the shadows in the front (opposite) side of the sphere are plotted by the solid (dashed)  lines. } 
\end{center}
\end{figure}

The three-dimentional description $(X, Y, Z)$ can be projected with the Lambert map projection (see, e.g. Qihe \cite{Q00}) given as 
\begin{eqnarray}
\left\{
	\begin{array}{l}
		\theta_x = k\cos u \sin v\\
		\theta_y = k\sin u
	\end{array}
\right.
~~~~~{\rm here}~~~
k=\sqrt{\frac{2}{1+\cos u\cos v}},~~~~{\rm and}~~~~
\left\{
	\begin{array}{l}
		u = \sin^{-1} Y\\
		v = \tan^{-1} (X/Z)
	\end{array}
\right.
.
\label{eq:Lambert}
\end{eqnarray}
In Fig. \ref{fig:bhsc_allskyL}, we plot the shapes of the shadows 
with the normalized Lambert map projection. The abscissa is $\theta_x$ and the
ordinate is $\theta_y$. In this projection given by Eq (\ref{eq:Lambert}), all the area of the
sky of the observer is mapped into the region of $\theta_x^2+\theta_y^2\le 1$. 
In this figure, the direction of the black hole ($\bar{\theta}=\pi$) 
corresponds to $(\theta_x,~\theta_y)=(0,~0)$, and the opposite direction 
($\bar{\theta}=0$) corresponds to the outermost circle (i.e., $\theta_x^2+\theta_y^2=1$) 
denoted by the dotted line. 
Both $\theta_x$ and $\theta_y$ are normalized so that  their maximum values 
become unity. The location of the observer is 
$\theta=85^\circ$ and $r=100,~10,~3,~2,~1.3,~1.1,~1.05,~1.045$. The
black hole spin is $a_*=0.999$. In this case, the radius of the event
horizon is $r_+\sim1.0447M$. In this figure, the outermost circle
denoted by the dotted line is the line of $\bar{\theta}=0$. As expected,
the size of the shadow becomes larger for the observer located closer to
the black hole. Interestingly, we can see that some photons in the
region of $\theta_x<0$, can escape to infinity even when the
observer is located very near the black hole. This is because of the
frame-dragging effect. It is confirmed that no photons emitted at the
horizon can escape. In Fig. \ref{fig:bhsc_allsky2} we plot 
the projection of the shadows onto the sky in the three dimensional
coordinate $(X,Y,Z)$, where the observer is located at the
origin, $(X, Y, Z)=(0, 0, 0)$ in this plot ({\it filled triangles}), and 
the direction of the black hole is $(X, Y,Z)=(0, 0, 1)$. While most of
the photons emitted from the location far 
from the hole (e.g. $r=100$) escape to infinity, many of the photons
emitted near the black hole are swallowed by the hole (e.g. $r=1.05$ and
1.045). The parameters are the same as in Fig. \ref{fig:bhsc_allskyL}. The
similar signatures seen in Fig. \ref{fig:bhsc_allskyL}can also be seen
in this figure. In Fig \ref{fig:bhsc_allsky2}, the shadow has a 
solid angle and its shape is deformed like in
Fig. \ref{fig:bhsc_allskyL}. In the next section, we give the new
equation for the solid angle.

\section{Solid angle for the photons trapped by the black hole}
\label{sec:trap}
Here we give the equation for the solid angle of the black hole shadow
in the sky of the LNRF observer. This solid angle is the function of the
black hole spin $a_*$ and the location of the observer $(r,
\theta)$. When the observer is moving with some velocity, the solid angle 
also depends on the velocity. Here we only consider the LNRF observer
for simplicity. Below we derive first the new equation for the solid angle of
the black hole shadow in the sky seen from an observer located at a
point near the black hole. Next, we show the distribution of the 
ratio of the solid angle in the sky in the $x$-$z$ plane.  Note that
this ratio corresponds to the ratio of the trapped photon by the hole
and the escape photon toward distant regions, when we consider the
photons isotropically emitted from a point (as a radiation source in LNRF)
near the black hole.  

\subsection{Equation for the solid angle }    

The shape of the black hole shadow is symmetric with respect to the
$Y$-$Z$ plane. This makes it sufficient to calculate the solid angles
surrounded by the $\theta_x$-axis and the contours of the black hole in
the region of $Y\ge 0$. The twice the amount of the solid angle for $Y\ge 0$ corresponds 
to the amount of the solid angle for the whole region of the black hole shadow. 
First, we consider the case when the whole region of the black hole shadow 
exists in the region of $Z\ge 0$, 
which is the front side of the observer when the observer faces the black hole. 
In this case, we have $s_r=-1$ and the equations describing the sphere
in $Z\ge 0$ are given as   
\begin{eqnarray}
Z=(1-X^2-Y^2)^{1/2}\equiv f^+(X,~Y),~~~~
f^+_X = \frac{-X}{(1-X^2-Y^2)^{1/2}},~~~~
f^+_Y = \frac{-Y}{(1-X^2-Y^2)^{1/2}}, 
\end{eqnarray}
where $f_X$ and $f_Y$ are the partial derivative of the function $f$ with respect to $X$ and $Y$, respectively. Now, the solid angle $S$ of the region $\Omega$ of the shadow on this sphere is calculated as 
\begin{eqnarray}
S=2\int \int_\Omega \sqrt{(f^+_X)^2+(f^+_Y)^2+1}~dX dY =2\int \int_\Omega \frac{dXdY}{\sqrt{1-X^2-Y^2}}.  	
\end{eqnarray}
Here we introduce the variables $(r,~\theta)$ as 
\begin{eqnarray}
X=r\cos\theta,~~~Y=r\sin\theta. 
\end{eqnarray}
From these definitions, we have 
\begin{eqnarray}
r=\sqrt{X^2+Y^2}=\sin\bar{\theta},~~~~
\tan\theta=\frac{Y}{X}=\frac{-1}{\tan\bar{\phi}},~~~~
\frac{\partial \theta}{\partial r_s}=\frac{\partial \bar{\phi}}{\partial r_s}. 
\end{eqnarray}
By using these, the solid angle $S$ is calculated as 
\begin{eqnarray}
S=2\int \int_\Omega =2\int \int_\Omega \frac{r}{\sqrt{1-r^2}}dr d\theta
=2\int_0^\pi d\theta \int_0^{r_{\rm max}(\theta)} \frac{r}{\sqrt{1-r^2}}dr
=2\int_0^\pi d\theta \left[1-\sqrt{1-r_{\rm max}^2(\theta)}\right]. ~~~~~
\end{eqnarray}
Because we now consider the case of $Z\ge 0$, i.e. $0\le \bar{\theta}\le \pi/2$, we obtain 
\begin{eqnarray}
\sqrt{1-r_{\rm max}^2(\theta)}=\sqrt{1-\sin^2\bar{\theta}}=\sqrt{\cos^2\bar{\theta}}=\cos\bar{\theta}. 
\end{eqnarray}
Next, by changing the integration variable from $\theta$ to $r_s$, we have 
the ranges for the integration as follows  
\begin{eqnarray}
  \theta=0\to \pi~{\rm and}~r_s=r_s^{\rm max} \to r_s^{\rm min},~~~~
  \mbox{~~~~ with ~~~~} 
  d\theta = \left(\frac{\partial \theta}{\partial r_s}\right) dr_s. 
\end{eqnarray}
Then, the solid angle $S$ is given as 
\begin{eqnarray}
S=2\int_{r_s^{\rm min}}^{r_s^{\rm max}}~dr_s~\left(-\frac{\partial \bar{\phi}}{\partial r_s}\right)
	(1-\cos\bar{\theta}), 
\end{eqnarray}
where 
\begin{eqnarray}
\cos\bar{\theta}=-s_r \left(\frac{1}{A_*}\right)^{1/2}\frac{\sqrt{R_*}}{1-\omega_*\zeta}~~~({\rm here}~~~s_r=-1). 
\end{eqnarray}

Now we we consider the case where the whole of the black hole shadow
exists in the region of $Z\le 0$ 
(the back side of the observer). 
In this case, we have $s_r=-1$ and the equations describing the sphere in $Z\le 0$ are given as 
\begin{eqnarray}
Z=-(1-X^2-Y^2)^{1/2}\equiv f^-(X,~Y),~~~~
f^-_X = \frac{X}{(1-X^2-Y^2)^{1/2}},~~~~
f^-_Y = \frac{Y}{(1-X^2-Y^2)^{1/2}}. 
\end{eqnarray}
In the same way, we have 
\begin{eqnarray}
\sqrt{1-r_{\rm max}^2(\theta)}=\sqrt{1-\sin^2\bar{\theta}}=\sqrt{\cos^2\bar{\theta}}=-\cos\bar{\theta}. 
\end{eqnarray}
The change of the sign of $f^\pm$ do not cause any change because both
$f^\pm_x$ and $f^\pm_y$ are used in the form of $(f^\pm_x)^2$ and
$(f^\pm_y)^2$, respectively. In terms of the change of the sign before
$\cos\bar{\theta}$, this change is canceled out by the change of the
sign of $s_r$. Consequently for the shadow in $Z<0$ 
the equation is essentially same as for the shadow in $Z\ge 0$. 
For the shadow crossing the $Z=0$, the equation assumes the same form. 

In summary, the equation of the solid angle $S$ is given by 
\begin{eqnarray}
 S=2\int_{r_s^{\rm min}}^{r_s^{\rm max}}~dr_s~\left(-\partial_{r_s}
      \bar{\phi}\right) (1-\cos\bar{\theta}), 
~~~~~~~{\rm where}~~~~~~~
\cos\bar{\theta}=-s_r
\left(\frac{1}{A_*}\right)^{1/2}\frac{\sqrt{R_*}}{1-\omega_*\zeta},~~~~~ 
\tan\bar{\phi}=\frac{s_\theta\rho_*^2}{A_*^{1/2}\sin\theta}
\left(\frac{\zeta}{\sqrt{\Theta_*}}\right).~~~~~~~
\label{eq:S}
\end{eqnarray}
Here the range of the integration for $r_s$ is determined by the condition of $\Theta_*\ge 0$, 
which can be numerically calculated. 
Moreover, by using the analytical form for $\zeta$, $\eta$, $\Theta_*$, $\bar{\phi}$, $\bar{\theta}$ and so on, the integrand of the integral of Eq. (\ref{eq:S}) is analytically given as 
\begin{eqnarray}
S&=&-4\rho^2 \sin\theta~\int_{r_s^{\rm min}}^{r_s^{\rm max}} 
	\Bigg[ 
	\frac{(r_s^3-3r_s^2+3r_s-a_*^2)[r_s^2(r_s-3)\zeta-a_*(r_s-1)(\eta+a_*^2\cos^2\theta)]}
		{a_*^2(r_s-1)^3(A\sin^2\theta\Theta_*+\rho^4\zeta^2)\sqrt{\Theta_*}}
	\left( A^{1/2}+\frac{s_r\sqrt{R_*}}{1-\omega_*\zeta} \right)
	\Bigg] dr_s. ~~~~~~~~
\label{eq:Sana} 
\end{eqnarray}
The derivation of Eq. (\ref{eq:Sana}) is given in App. \ref{app:Sanaderiv}. The Eqs. (\ref{eq:S}) and (\ref{eq:Sana}) are the main results of the present study. 

\subsection{ Photon trapped ratio }  

\begin{figure}
\begin{center}
\vspace{5mm}
\hspace*{-5mm}
\hspace{3.5mm}
\includegraphics[width=10mm, angle=270]{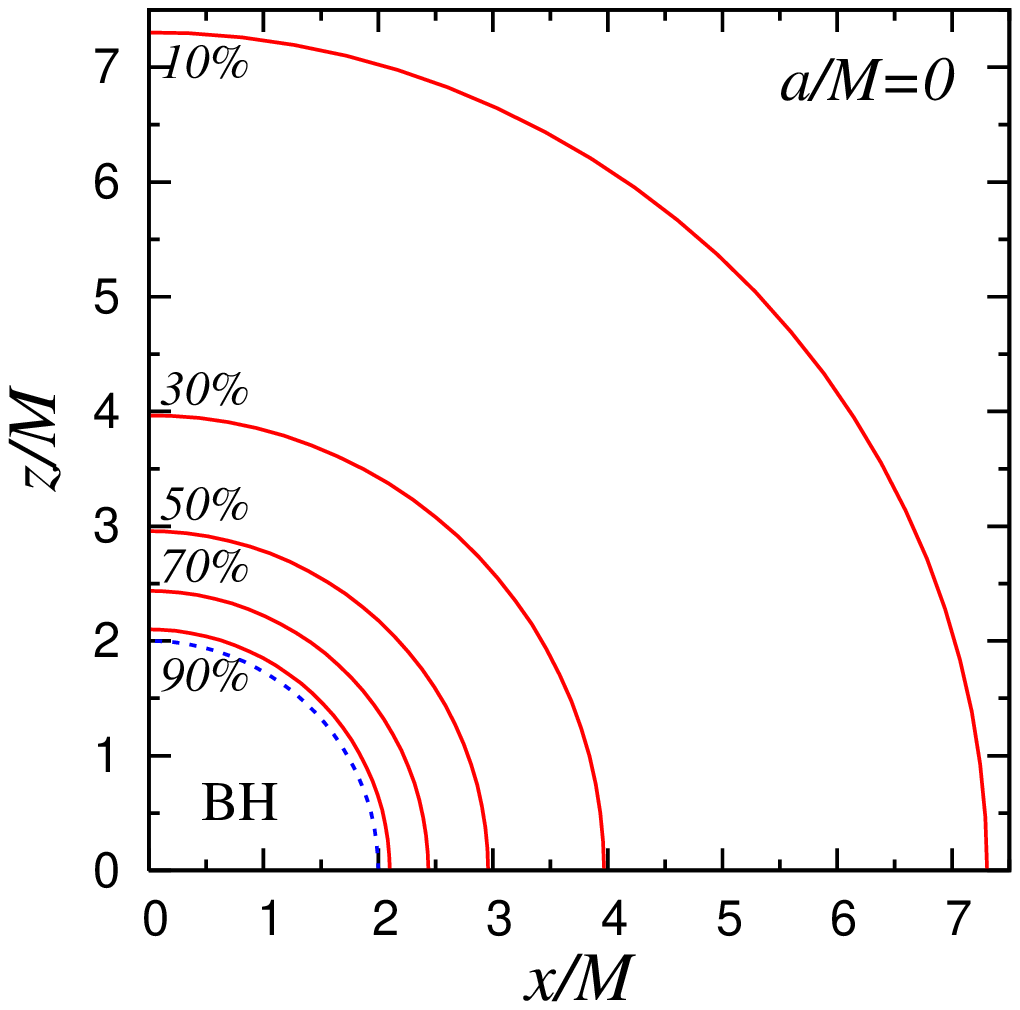}\vspace{80mm}\\ 
\includegraphics[width=10mm, angle=270]{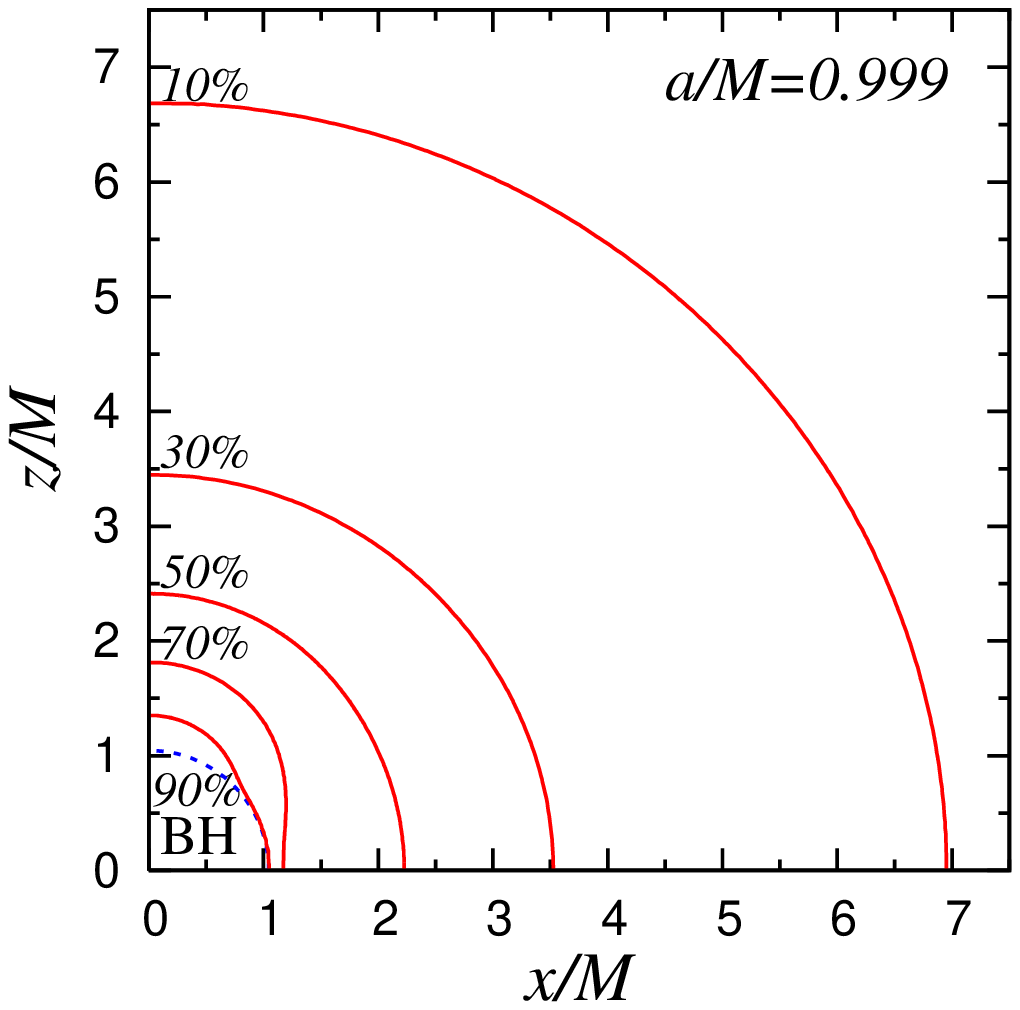}\vspace{0mm}
\vspace{75mm}
\caption{\label{fig:trap}
Photon trapped ratios $S/(4\pi)$ around the black holes with spins of $a_*=0$ ({\it top}) and 0.999 ({\it bottom}). We draw the contours of the trapped ratios for 10\%,~30\%,~50\%,~70\%,~90\% ({\it solid lines}) and the event horizon ({\it dotted lines}). }
\end{center}
\end{figure}

\begin{figure}
\begin{center}
\vspace{5mm}
\hspace*{-5mm}
\hspace{3.5mm}
\includegraphics[width=10mm, angle=270]{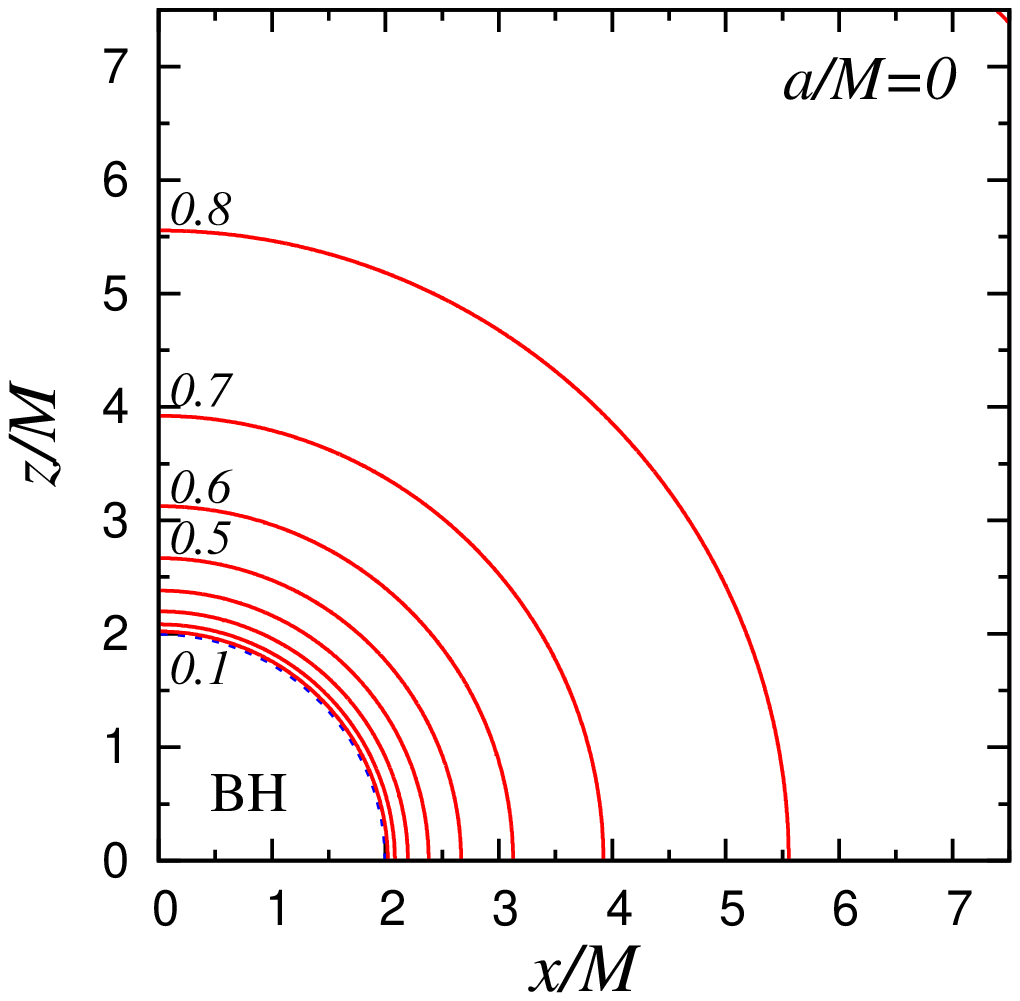}\vspace{80mm}\\
\includegraphics[width=10mm, angle=270]{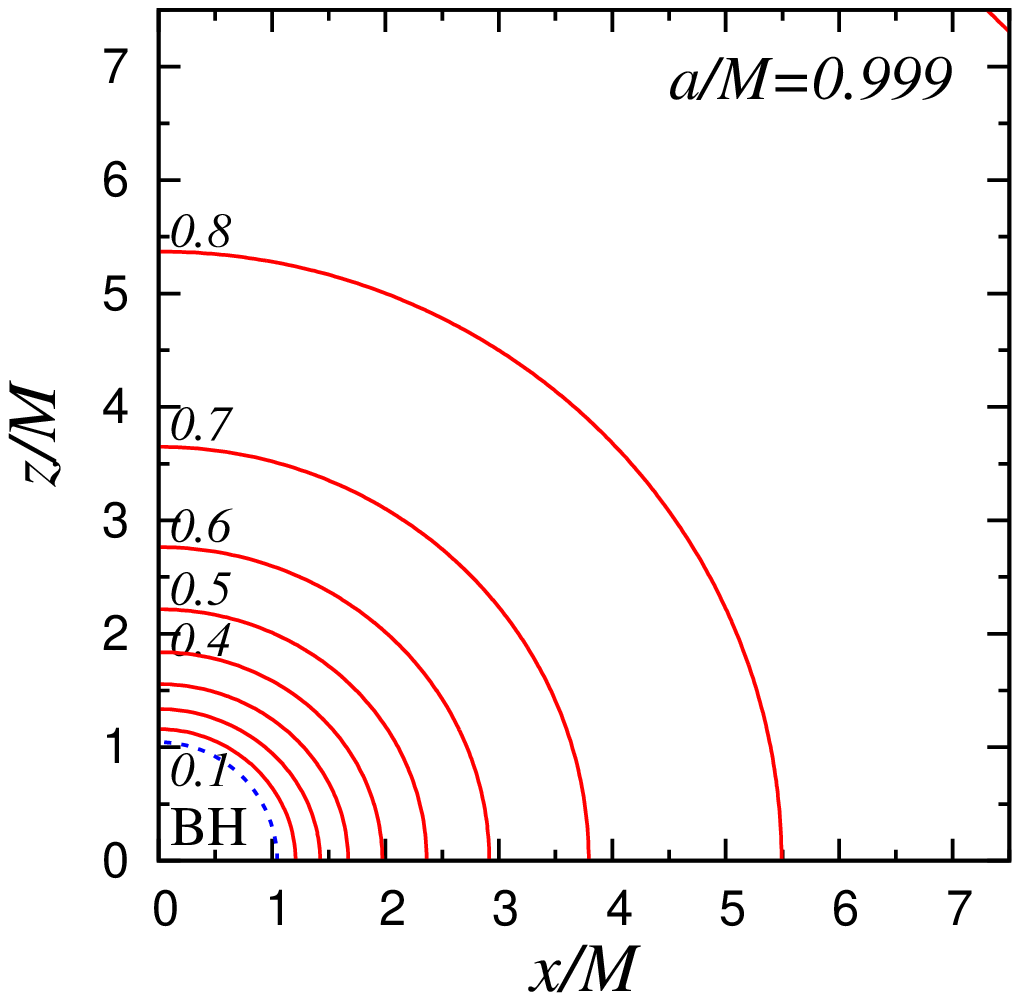}\vspace{0mm}
\vspace{75mm}
\caption{\label{fig:gr}
Lapse function $\alpha$ representing effects of the gravitational redshift for the black holes  
with spins of $a_*=0$ ({\it top}) and 0.999 ({\it bottom}). We draw the contours of the lapse function for $\alpha=$0.1,~0.2,~0.3,~0.4,~0.5,~0.6,~0.7,~0.8,~0.9 ({\it solid lines}) and the event horizon ({\it dotted lines}). We can see the contours for $\alpha=0.9$ around the regions near $(x/M, z/M)=(0.75,~0.75)$ in both panels. At the event horizon the value of $\alpha$ becomes zero. }
\end{center}
\end{figure}

Based on Eqs. (\ref{eq:S}) and (\ref{eq:Sana}), we numerically calculate the distribution of the 
photon trapped ratio $S/(4\pi)$ around the black hole. In Fig.\ref{fig:trap} we plot the trapped ratios 
$S/(4\pi)$ for $a_*=0$ ({\it top}) and 0.999 ({\it bottom}) in the $x$-$z$ plane. For the rapidly rotating 
black hole, the photon trapping ratio decreases especially near the equatorial plane because of the 
frame-dragging effects. That is, the photons near the equatorial plane can escape to infinity 
more easily than the photons near the pole. These features can also be seen in the shapes of the 
shadows. Surprisingly, as shown in these plots, 10 \% to 30 \% of the photons emitted very close 
to the black hole (e.g. $r<1.2$ for $a_*=0.999$) can escape to infinity. Although these 
calculations do not take into account the effects of the gravitational redshift, these escaping 
photons can transport the information of the spacetime and the general relativistic phenomena to 
the observer at infinity. These results about the trapped ratio for photons are very important 
when we consider the radiative phenomena that arise near the event horizon. In Fig. \ref{fig:gr} we plot   
the lapse function $\alpha$ representing the effects of gravitational redshift for black hole 
spins of $a/M=0$ ({\it top}) and 0.9 ({\it bottom}). The contours of the lapse function in Fig. \ref{fig:gr} 
shows nearly spherical shapes, which is contrast to the plots of the photon trapped ratios in 
Fig. \ref{fig:trap}. Then, even if we take the effects of the gravitational redshift into account, the 
energy amount of the escaping photons near the equatorial plane is  than that near the pole 
region for the case of the rapidly rotating black hole. 

\section{Discussion and conclusions}
\label{sec:con}
Since the last century, observational data containing information of the black hole 
spacetime have been obtained for the black hole candidates in e.g. the active galactic nuclei, 
the X-ray black-hole binaries and the Galactic center of our galaxy. However, the parameters 
of the black hole spacetime like mass and spin estimated in the past studies usually and 
highly depend on the assumed accretion flow models, 
and so far the final values of the parameters of 
the black hole, especially the black hole spin, have not been obtained for any black hole candidates. 
This is because the observational data usually contain the information of the plasma physics of 
the accreting matter in addition to the information of the spacetime metric, and frequently both these 
informations (metric and plasma physics) are degenerate in the observational data. 
In these cases we cannot uniquely determine the spacetime metric from the observational data. 
Because the optical phenomena directly reflect the spacetime metric, 
the spacetime metric can be uniquely determined if the observational signatures closely related to 
the optical phenomena like the contours of the black hole shadow can be found in the 
observational data. For this purpose a deep understanding of the optical phenomena around the 
black hole is important. 

In past studies it was proposed that the 
black hole's metric can be estimated from the observational features like the line and 
continuum spectrum, quasi-periodic oscillation (QPO), 
radio visibility, polarization 
(see, reviews, e.g., Remillard \& McClintock \cite{RM06}; Psaltis \cite{P08}). 
However, all these observational signatures highly depend on the accretion flow models.   
Actually, for some black hole candidates, several values of the black hole spin can be 
proposed for one and the same black hole candidate. 
For example, in terms of the iron line spectrum observed 
in an X-ray black hole binary, it is pointed out that the Fe K line profile and the estimated value of the 
black hole spin of the black hole binary GX 339-4 highly depend on the assumed 
continuum spectrum, which is first subtracted from the observational data to make the line profile 
(Yamada et al. \cite{Y09}). This means that the black hole spin cannot be correctly estimated from 
the line profile unless we can uniquely determine the accretion flow model, which produces  
the continuum spectrum. On the other hand, in terms of the observed QPO phenomena, 
the origin of the QPO is still under debate, although some theoretical models of the QPO like 
the disk oscillation model have been proposed. In terms of X-ray polarization of the accretion 
disk, the observed polarization features highly depend on the corona models. However, we do not 
fully understand the production mechanism and physical properties like the spatial 
distribution of the corona. Therefore, to understand the observational signatures 
containing the information of the spacetime, the physics of the accretion flows and outflows 
(and/or jets) around the black holes should be deeply investigated and understood especially 
by means of magnetohydrodynamic simulations with effects of the radiation fields. 
Actually, recently magnetohydrodynamic simulations with effects of the radiation are attempted 
(e.g. De Villiers \cite{V08}, Farris, Liu \& Shapiro \cite{F08}, Mo\'{s}cibrodzka et al. \cite{M09}). 
In the near future, the radiation fields around the black hole spacetime will be exactly solved by 
the dynamical simulations with effects of the anisotropy of the radiation field considered in this 
paper. Because the equation given in this study provide the exact results of the contours of the black 
hole shadow and the photon trapped ratio, the equation given in this study can be used for the 
code tests in the developments of simulation code of the radiation field.  

We have the appearance of the black hole
(the black hole shadow) seen by the observer located near the black hole. 
In summary the calculation method for the photon trapped ratio presented in this study 
can be extended to the astropysically more plausible cases.  Although in this study 
we assume the LNRF, the calculation in this study can be extended to other reference 
frames moving with astrophysically frequently considered velocity fields like 
corotating or counterrotating Keplerian disks. 
Actually, in the past studies, other frames like a static frame, a circular geodesic frame 
(or a Keplerian frame) and a radially falling frame considered (e.g. Takahashi \cite{T07a}; 
Schee \& Stuchl\'{i}k \cite{SS09a,SS09b}) in addition to the LNRF. 
Especially Schee \& Stuchl\'{i}k (\cite{SS09a}) 
gave the equation and contours of the black hole silhouettes by using these frames. 
In addition to the contours, the equation of the photon trapping ratio presented in this 
study [Eqs. (\ref{eq:S}) and (\ref{eq:Sana})] can be straightforwardly extended to 
a more general velocity field. 

\begin{acknowledgements}
RT is grateful to Professors K. Makishima, Y. Eriguchi and S. Mineshige
for their continuous encouragements, and to K. Ohsuga, Y. Kato, 
J. Fukue, Y. Sekiguchi for useful discussion. The authors thank the referee, Prof. Z. Stuchl\'{i}k, 
for valuable comments and suggestions, which improved the original manuscript. 
This research is supported by the Grant-in-Aid for Scientific Research Fund of the 
Ministry of Education, Culture, Sports, Science and Technology, Japan 
[Young Scientists (B) 21740149 (RT)]. 
\end{acknowledgements}

\appendix
\section{Derivation of the equation (\ref{eq:Sana})}
\label{app:Sanaderiv}
First,  we give the analytic expression for the partial derivative $\partial\bar{\phi}/\partial r_s$ appeared in Eq. (\ref{eq:S}). 
Both $\eta$ and $\zeta$ are the function of $r_s$ calculated as 
\begin{eqnarray}
\zeta = \frac{-1}{a_*(r_s-1)}(r_s^3-3r_s^2+a_*^2r_s+a_*^2),~~~
\eta = \frac{-r_s^3}{a_*^2(r_s-1)^2}(r_s^3-6r_s^2+9r_s-4a_*^2). 
\end{eqnarray}
Based on this, the derivatives of these with respect to $r_s$ is analytically calculated as 
\begin{eqnarray}
(\partial_{r_s}\zeta)=\frac{-2}{a_*(r_s-1)^2}(r_s^3-3r_s^2+3r_s-a_*^2),~~~~
(\partial_{r_s}\eta)=\frac{-4r_s^2(r_s-3)}{a_*^2(r_s-1)^3}(r_s^3-3r_s^2+3r_s-a_*^2). 
\label{eq:zetaetaA}
\end{eqnarray}
From Eq. (\ref{eq:zetaetaA}), we know the relation of 
\begin{eqnarray}
(\partial_{r_s}\eta)=\frac{2r_s^2(r_s-3)}{a_*(r_s-1)}(\partial_{r_s}\zeta). 
\end{eqnarray}
On the other hand, the partial derivative of $\Theta_*$ with respect to $r_s$ is given as 
\begin{eqnarray}
(\partial_{r_s}\Theta_*) = -\frac{2\zeta}{\tan^2\theta}(\partial_{r_s}\zeta)+(\partial_{r_s}\eta),~~~~
(\partial_{r_s}\sqrt{\Theta_*}) = \frac{\partial_{r_s}\Theta_* }{2\sqrt{\Theta_*}},
\end{eqnarray}
If we introduce 
\begin{eqnarray}
n^{(x)} \equiv \frac{p^{(\theta)}}{p^{(t)}}= \sin\bar{\theta}\cos\bar{\phi},~~~~
n^{(y)} \equiv \frac{p^{(\phi)}}{p^{(t)}}= \sin\bar{\theta}\sin\bar{\phi},~~~~
n^{(z)} \equiv -\frac{p^{(r)}}{p^{(t)}}=\cos\bar{\theta},
\end{eqnarray}
and 
\begin{eqnarray}
n^{(w)} \equiv \frac{n^{(y)}}{n^{(x)}} = \tan\bar{\phi}=\frac{s_\theta\rho^2}{A^{1/2}\sin\theta}\left(\frac{\zeta}{\sqrt{\Theta}}\right),
\end{eqnarray}
we have 
\begin{eqnarray}
\partial_{r_s}\bar{\phi} &=& \cos^2\bar{\phi}~(\partial_{r_s}n^{(w)}) \\
	&=& \frac{1}{1+(n^w)^2} \left(\frac{s_\theta\rho^2}{A^{1/2}\sin\theta}\right)
		\partial_{r_s}\left(\frac{\zeta}{\sqrt{\Theta}}\right)\\
	&=& \frac{s_\theta A^{1/2}\rho^2 \sin\theta}{2\sqrt{\Theta_*}(A\sin^2\theta\Theta_*+\rho^4\zeta^2)}
		\left[ 2(\eta+a_*^2\cos^2\theta)(\partial_{r_s}\zeta)-\zeta(\partial_{r_s}\eta) \right]. 
\end{eqnarray}
On the other hand, the factor $1-\cos\bar{\theta}$ in the integral given in Eq. (\ref{eq:S}) is calculated as 
\begin{eqnarray}
1-\cos\bar{\theta} = 1+\left(\frac{s_r}{A^{1/2}}\right)\frac{\sqrt{R_*}}{1-\omega_*\zeta}. 
\end{eqnarray}
By using the analytic expressions for $\partial_{r_s}\bar{\phi}$ and $1-\cos\bar{\theta} $, the solid angle $S$ is calculated as 
\begin{eqnarray}
S&=&2\int_{r_s^{\rm min}}^{r_s^{\rm max}} dr_s 
	\frac{A^{1/2}\rho^2 \sin\theta}{2\sqrt{\Theta_*}(A\sin^2\theta\Theta_*+\rho^4\zeta^2)}
		\left[ \zeta(\partial_{r_s}\eta)-2(\eta+a_*^2\cos^2\theta)(\partial_{r_s}\zeta) \right] 
	\left[ 1+\left(\frac{s_r}{A^{1/2}}\right)\frac{\sqrt{R_*}}{1-\omega_*\zeta} \right] \\	
	&=&2\rho^2 \sin\theta~\int_{r_s^{\rm min}}^{r_s^{\rm max}} dr_s 
	\frac{(\partial_{r_s}\zeta)}{\sqrt{\Theta_*}(A\sin^2\theta\Theta_*+\rho^4\zeta^2)}
		\left[ \frac{r_s^2(r_s-3)}{a_*(r_s-1)}\zeta-\eta-a_*^2\cos^2\theta \right] 
	\left( A^{1/2}+\frac{s_r\sqrt{R_*}}{1-\omega_*\zeta} \right) \\
	&=&-4\rho^2 \sin\theta~\int_{r_s^{\rm min}}^{r_s^{\rm max}} dr_s 
	\frac{(r_s^3-3r_s^2+3r_s-a_*^2)[r_s^2(r_s-3)\zeta-a_*(r_s-1)(\eta+a_*^2\cos^2\theta)]}
		{a_*^2(r_s-1)^3(A\sin^2\theta\Theta_*+\rho^4\zeta^2)\sqrt{\Theta_*}}
		\left( A^{1/2}+\frac{s_r\sqrt{R_*}}{1-\omega_*\zeta} \right), ~~~~~~~~~~~~~~
\end{eqnarray}
where the analytic expressions for $\zeta$, $\eta$,  $\Theta_*$ and $R_*$ are given above. 

\end{document}